\documentclass{JHEP3}
\usepackage{amssymb}
\usepackage{amsmath}
\usepackage{extarrows}
\usepackage{epsfig}

\newcommand\ee{\end{equation}}
\newcommand\be{\begin{equation}}
\newcommand\eea{\end{eqnarray}}
\newcommand\bea{\begin{eqnarray}}

\def\l{\left}
\def\r{\right}
\def\beq{\begin{equation}}
\def\eeq{\end{equation}}
\def\de{\partial}

\def\ov{\overline}
\def\nn{\nonumber}
\def\barr{\begin{array}}
\def\earr{\end{array}}

\def\cF{{\cal F}}
\def\cK{{\cal K}}
\def\ie{{\it i.e.}}
\def\Re{{\rm Re}}

\def\D{{\cal D}}

\title{On general flux backgrounds with localized sources}
\author{
Giovanni Villadoro \\ 
Jefferson Laboratory of Physics, Harvard University, \\
Cambridge, Massachusetts 02138, USA \\
E-mail: \email{villador@physics.harvard.edu}}
\author{
Fabio Zwirner \\
Dipartimento di Fisica, Universit\`a di Padova and INFN, \\ 
Sezione di Padova, Via Marzolo 8, I-35131 Padova, Italy \\
E-mail: \email{fabio.zwirner@pd.infn.it}}
\preprint{DFPD-07/TH/17}
\abstract{
We derive new consistency conditions for string compactifications with generic fluxes (RR, NSNS, geometrical) and localized sources (D-branes, NS-branes, KK-monopoles). The constraints are all related by string dualities and share a common origin in M-theory. We also find new sources of instabilities. We discuss the importance of these conditions for the consistency of the effective action and for the study of interpolating solutions between vacua.}

\begin{document}
\section{Introduction}

The search for realistic string compactifications can rely today on an impressive theoretical toolbox (for recent reviews and references to the original literature, see e.g. \cite{revgrana, revkachru, revlust}). Two very important ingredients, whose understanding is steadily progressing, are {\it fluxes} and {\it localized sources}. In their different manifestations, connected by string dualities, they play a r\^ole in all ten-dimensional string theories and also in M-theory, whose field-theory limit is eleven-dimensional supergravity. 

As is well known, fluxes are the (quantized) cohomologically non-trivial parts of $p$-form field strengths in the compact internal space. In each ten-dimensional string theory, or its effective ten-dimensional supergravity, we can consider general systems of fluxes associated to the available classical degrees of freedom\footnote{The various string dualities, and their counterparts in the effective supergravities, suggest the existence of more general  `non-geometric' fluxes: we will neglect here, for simplicity, this interesting generalization. For a recent review see \cite{wecht}.}. For example, in the Neveu--Schwarz--Neveu--Schwarz (NSNS) sector of type-II theories and in the heterotic theory, we can consider fluxes of the 3-form field strength $H$. In type-II theories, we can consider fluxes of the different Ramond--Ramond (RR) $p$-form field strengths  $G^{(p)}$ ($p$ even for type-IIA, odd for type-IIB). In the heterotic theory, we can consider magnetic fluxes associated with the field strength $F$ of the $E_8 \times E_8$ or $SO(32)$ gauge group. In type-II theories, we can consider magnetic fluxes associated with the gauge field strengths $\cF$  localized on D-branes. Finally, in all ten-dimensional string theories we can consider the so-called geometric (or metric) fluxes, associated with the internal components of the spin connection $\omega$ (suitably anti-symmetrized and with curved-space indices), equivalent to coordinate-dependent compactifications with a Scherk--Schwarz twist. This classification can be extended to M-theory, or its effective eleven-dimensional supergravity: in such a case the available fluxes are those of the 4-form $F^{(4)}$, plus geometric fluxes and possible fluxes for localized gauge field strengths on M-branes.

Fluxes are in general accompanied by localized sources, extending over a submanifold of the whole ten-dimensional (eleven-dimensional for M-theory) space-time. For example, fundamental strings (NS1-branes) and NS5-branes are the electric and magnetic sources of the NSNS 2-form potential $B$ and its dual. Analogously, D$p$-branes are the electric and magnetic sources for the RR $(p+1)$-form potentials $C^{(p+1)}$ and their duals. KK-gravitons and KK5-monopoles are the electric and magnetic sources for geometric fluxes in ten dimensions. In M-theory, M2-branes and M5-branes source the 3-form $A^{(3)}$ and its dual 6-form, whilst the magnetic source for geometrical fluxes are KK6-monopoles. In all theories, we may also need to include non-dynamical objects such as orbifold/orientifold planes in the general set of localized sources. 

The aim of this paper is to provide a unified and generalized description of the consistency conditions for string compactifications with fluxes and localized sources, extending the presently known results. In particular, we will concentrate on the so-called {\it localized} Bianchi Identities (BI), which enforce on the effective action those non-anomalous local invariances of the underlying compactified theory that are associated with localized gauge fields. Their r\^ole is similar, but still less generally understood, to the r\^ole of the {\it bulk} BI, associated with the local symmetries of the metric and of the $p$-form potentials  
in the NSNS and RR sectors. The integrability conditions of the bulk and localized BI constrain the allowed combinations of fluxes and localized sources, and play a crucial r\^ole in enforcing the local symmetries (gauge invariance, supersymmetry) in the effective four-dimensional theory. These conditions are particularly important when we cannot get a full-fledged solution of the higher-dimensional equations of motion, and we must rely on an effective field theory approach to explore the consequences of different systems of fluxes and localized sources, making sure that the local symmetries of the underlying theory are duly respected. 

The best known localized BI is the one associated with the Freed--Witten (FW) anomaly \cite{FW}, which provides a constraint on the flux $\ov{H}$, i.e. the cohomologically non-trivial part of the NSNS 3-form $H$, in the presence of D$p$-branes: 
$$
\ov{H} \wedge [\pi_p]= 0
\qquad \Leftrightarrow \qquad 
\int_{\gamma_3 \subset \pi_p} H = 0 \,,
$$
where $\pi_p$ denotes the D$p$-brane world-volume, $[\pi_p]$ the $(9-p)$-form Poincar\'e dual to it and $\gamma_p$ is a generic $p$-cycle. It was recently shown, on the basis of complementary arguments \cite{CFI,VZD}, that a more general localized BI holds in the presence of non-trivial geometrical fluxes $\ov{\omega}$:
$$
\ov{\omega} \, [\pi_p] = 0 \, , 
\qquad
\ov{H} + \ov{\omega} \, \ov{\cF} = 0 \, ,
$$
where the overline has the same meaning as before. The FW condition plays also a crucial r\^ole in the attempts at embedding the classification of brane charges within K-theory, see e.g. \cite{Evslinrev} for a review and references to the original literature. Following this approach interesting results have been obtained recently in \cite{Ma,EM,KM}. However, it seems that K-theory is not general enough to give a complete description of branes and flux charges. Indeed, K-theory seems to be incompatible with S duality \cite{DMW}. In this paper we will adopt instead an effective field theory approach, exploiting the power of the supergravity description. We will be able to further generalize the above constraints to others that involve RR fluxes on NS5-branes or KK5-monopoles,
$$
[\nu_5] \wedge \ov G = 0
\qquad \Leftrightarrow \qquad 
\int_{\gamma_p  \subset \nu_5} G^{(p)} = 0 \,,
$$
and 
$$
[\kappa_5]^\xi \ \ov G_\xi=0\qquad \Leftrightarrow \qquad 
\int_{\gamma_p \subset \kappa_5} G_\xi^{(p+1)} = 0 \,,
$$
where $\xi$ is the fibered circle of the KK5-monopole and more details on the notation will be given later in the text. We will also show that a similar condition holds for RR fluxes on D$p$-branes:
$$
\int_{\gamma_p \subset \pi_p} G^{(p)} = 0 \, .
$$

We have already stressed that all the above localized BI are instrumental for enforcing the local symmetries of the effective four-dimensional theory. The generic effect of fluxes is to gauge some axionic symmetries, whereas Euclidean (instantonic) branes generically induce in the effective action terms that break some of these axionic symmetries. As already realized in some special cases (see, e.g., \cite{KT,AZ,BK}), and as we will discuss in more general terms in this paper, localized BI prevent this clash from happening.

The previous considerations on localized BI were implicitly assuming free (albeit possibly intersecting) branes. It is however known \cite{MMS} that the FW anomaly-cancellation condition can be relaxed in the presence of branes ending on the `anomalous' brane, and that this may lead to instabilities such as the decay of $N$ D$(p-2)$-branes into $N$ units of  $H$-flux via an instantonic D$p$-brane. In this paper we generalize our new localized BI in a similar way, and identify the new corresponding instabilities that arise. We also study the r\^ole of these configurations when dealing with domain walls interpolating between two string vacua.

The rest of this paper is organized as follows. In Section~2 we first recall the localized BI associated with the Freed--Witten (FW) anomaly, which provides a constraint on the NSNS 3-form flux $\ov{H}$, and review how by T-duality we can derive a similar localized BI involving also the geometrical flux $\ov{\omega}$. Then we make use of S-duality to infer the existence of additional new localized BI, valid for NSNS localized sources (NS5-branes and KK5-monopoles) and constraining RR fluxes. In Section~3 we show how all the above localized BI have a common origin from M-theory. We also provide some examples of how the new localized BI are essential for ensuring gauge invariance of the effective action, which could be spoiled by the simultaneous presence of fluxes and Euclidean branes inducing instanton effects. In Section~4 we generalize our considerations to the case of branes ending on branes. We discuss the new sources of instabilities that arise from the new allowed configurations of branes and fluxes. We stress the importance of these configurations in connecting string vacua with different branes and fluxes (and geometry as well). In Section~5 we provide the full derivation of all constraints directly in 10D, using the bulk BI. In the final section we summarize our conclusions. 

\section{D-branes on flux backgrounds}
\label{sec:rev}

In type-II string theories, D$p$-branes are the electric and magnetic sources of the Ramond--Ramond (RR) $(p+1)$-form potentials $C^{(p+1)}$ and their duals, in the same way as fundamental strings and NS5-branes are the electric and magnetic sources of the Neveu--Schwarz--Neveu--Schwarz (NSNS) 2-form potential $B$ and its dual. In the effective field theory, this fact is manifest in the Wess--Zumino (WZ) part of the D$p$-brane action,
\beq 
\label{eq:SWZ1}
S_{WZ}=\int_{\pi_p} C \, e^{\cF} \, ,
\eeq
where $\pi_{p}$ is the $(p+1)$-dimensional world-volume of the brane, $C$ is a convenient way of combining all the RR potentials into a single polyform, and $\cF = F-B$ is the brane-localized gauge field. If we assume the ten-dimensional space-time to be of the form ${\mathbb R}\times X_9$, where $X_9$ is the nine-dimensional manifold associated with the space coordinates, then $\pi_p$ may extend along the time direction, wrapping a $p$-dimensional sub-manifold of $X_9$, or be localized in time, wrapping a $(p+1)$-dimensional sub-manifold of $X_9$: in the latter case the D$p$-brane is also called an instantonic or Euclidean D$p$-brane. Since the Dirac--Born--Infeld (DBI) part of the brane action is proportional to its volume, the brane wants to shrink. Therefore, to be stable the brane needs to wrap  a non-contractible cycle. For this reason brane charges are usually classified by homology. However, at variance with the usual minimal couplings between sources and gauge fields, the brane action depends also on the localized gauge field $\cF$,  which encodes the tangent\footnote{The normal ones, which describe the position of the branes, arise instead from the pull-back of the bulk fields on the branes, which we keep implicit in our notation.} degrees of freedom of the open strings ending on the D$p$-branes. Because of this a D$p$-brane does not couple only to the RR $(p+1)$-form, but also to lower-rank potentials. Indeed, eq.~(\ref{eq:SWZ1}) reads, more explicitly:
\beq 
\label{eq:SWZexp}
S_{WZ}=\int_{\pi_p} \l [ C^{(p+1)} + C^{(p-1)}\wedge \cF + \frac12 \, C^{(p-3)}\wedge \cF\wedge \cF + \dots \r] \, .
\eeq
To account for this effect,  K-theory has been proposed to classify D-brane charges instead of homology \cite{MM,W}.

If $\pi_p$ is the $(p+1)$-dimensional sub-manifold of ${\mathbb R} \times X_9$ occupied by the brane world-volume, we call $[\pi_p]$ the $(9-p)$-form Poincar\'e dual to $\pi_p$. The requirement that $\pi_p$ is a cycle implies, via Stokes theorem, that
\beq 
\label{eq:WI1}
\de \pi_p=0 \qquad \Leftrightarrow \qquad d[\pi_p]=0\,,
\eeq
which is the generalization of the Ward identity to extended objects. As mentioned before, the DBI part of the brane action is proportional to its volume, so that stability requires the cycle to be non-trivial, \ie
\beq 
\label{eq:stab1}
\pi_p\neq \de \pi'_{(p+1)}
\qquad \Leftrightarrow \qquad
[\pi_p]\neq d[\pi'_{(p+1)}]\, ,
\eeq
for every $(p+1)$-cycle $\pi'_{(p+1)}$. 

However, the D-brane ``current" is modified by the presence of the localized gauge field $\cF$, and eq.~(\ref{eq:WI1}) becomes
\beq 
\label{eq:WI2}
d\l ( [\pi_p]e^\cF\r)=0\,.
\eeq  
This is expected when looking at the Bianchi Identities (BI) for the RR gauge fields. Since D$p$-branes couple electrically to the RR $(p+1)$-form potentials $C^{(p+1)}$ [eq.~(\ref{eq:SWZexp})], they act as magnetic sources for the dual fields $C^{(7-p)}$, and appear in the BI of the $(8-p)$-form RR field strengths\footnote{Notice that we adopt here the dual formulation for the RR forms, not the democratic one, following the conventions of refs.~\cite{BKORV,VZ2A}.} $G^{(8-p)}$:  
\beq 
\label{eq:BI1}
d G^{(8-p)} + H\wedge G^{(6-p)}=\sum_q Q_{p}(\pi_q) \, ,
\eeq
where $H$ is the 3-form NSNS field strength and $Q_{p}(\pi_q)$ is the D$p$-brane charge (density) of the D$q$-brane wrapping the cycle $\pi_q$, \ie\ the projection over $(9-p)$-forms of the D$q$-brane charge $[\pi_q]e^\cF$. In the absence of $H$, the second term on the l.h.s. of eq.~(\ref{eq:BI1}) vanishes
and eq.~(\ref{eq:WI2}) follows from the closure of the external derivative ($dd=0$).

We can understand the meaning of eq.~(\ref{eq:WI2}) by expanding the exponential. By projecting over $p$-forms we get two conditions:
\beq
d[\pi_p] = 0 \, ,
\qquad
d\cF = 0 \, .
\label{eq:cWI2}
\eeq
The first is the usual homology condition (\ref{eq:WI1}), the second is nothing else than the localized BI for the gauge fields on the D$p$-branes.

However, eq.~(\ref{eq:WI2}) is not yet complete. For non-trivial $H$ fields, we get an extra term in eq.~(\ref{eq:WI2}), an extra source. It is a general fact that, in type-II string theory, external derivatives are always dressed with an $H$ contribution. If we define the modified external derivative
\beq
d_H\equiv d+H\wedge\,, \nn
\eeq
then we can rewrite the RR BI of eq.~(\ref{eq:BI1}) in the short-hand notation
\beq 
\label{eq:RRBI1}
d_H G=Q_{R} \, .
\eeq
There is no real advantage in this notation as long as we do not define some useful property for the modified derivative operator $d_H$. In the absence of magnetic sources for the NSNS 2-form (\ie\ NS5-branes), the BI for $H$,
\beq
dH=0 \, , \nn
\eeq
implies
\beq
d_H d_H =0\,. \nn
\eeq
Since, as it is the case for the usual external derivative, $d_H$ is closed,  we can try to construct a ``twisted" cohomology starting from this modified derivative.  The corresponding modified classification 
for D-brane charges is usually called {\it twisted} K-theory \cite{twk}. We can now use the closure of $d_H$ and act with it on both sides of eq.~(\ref{eq:RRBI1}). This gives the condition
\beq
d_H \l ( [\pi_p]e^\cF\r)=0\,. \nn
\eeq
By expanding this equation we recover the first condition of eq.~(\ref{eq:cWI2}), while
the second one gets modified into\footnote{For simplicity, here and in the following we will consider
trivial background values for the localized field strenghts ($\cF$ for D$p$-branes, $\cK$ for NS5-branes, etc.). 
The latter can easily be included by taking into account their contributions
to the solutions of the Bianchi identities for the NSNS and RR forms.}
\beq 
\label{eq:lBI}
d_H \cF=d\cF +H=0\,.
\eeq
It can be easily checked that this is the correct modified BI for the D-brane gauge fields, by solving for $\cF$. In the absence of fluxes the result reads
\beq
\cF=d A - B=F-B\,, \nn
\eeq
which is the correct (gauge-invariant) combination entering the D-brane action. Therefore the NSNS field $H$ is a magnetic source for the localized gauge fields $\cF$ on the D-brane.  Since the first term in eq.~(\ref{eq:lBI}) is a total derivative, $H$ calculated on the D-brane world-volume must be trivial in cohomology, \ie
\beq 
\label{eq:FW}
\ov{H} \wedge [\pi_p]= 0
\qquad \Leftrightarrow \qquad 
\int_{\gamma_3 \subset \pi_p}H=0\,,
\eeq
which corresponds to the  cancellation of the Freed--Witten (FW) anomaly\footnote{The actual constraint
reads $\l.\ov H\r|_{\pi_p}+W_3(\pi_p)=0$, but we will restrict ourselves to the cases where $\pi_p$ is a $Spin^c$ manifold, whose third Stiefel--Whitney class $W_3(\pi_p)$ vanishes. See \cite{Sparks} for a discussion of this more general case in a context related to ours.}~\cite{FW}. Here and in the following, we denote with $\ov{H}$ the cohomologically non-trivial part of $H$ (similarly for other fluxes\footnote{As for fluxes also for localized sources one should distinguish between the local contribution to the BI and the associated cohomological charge. To keep notation light we use the same notation for the two cases as it should be clear from the context which of the two applies.}), and with $\gamma_p$ a generic $p$-cycle. The constraint of eq.~(\ref{eq:FW}) can be relaxed only if other sources for the localized gauge fields are added. This is the case \cite{Str,Tbis} when other D($p-2$)-branes {\it end} on $\pi_p$. As shown in \cite{MMS}, this also allows normal D$p$-branes to decay into instantonic D($p+2$)-branes supporting an $H$ flux. In the following we will focus our attention on free brane configurations, where branes can intersect but not end on each other. We will return on the interesting case of branes ending on branes later, in section~\ref{sec:MMS}.

The importance of the RR BI in type-II string compactifications is well known. Integrating eq.~(\ref{eq:BI1}) on a compact space gives a highly non-trivial  constraint---Gauss law implies the cancellation of the total charge. This condition is crucial for the effective theory to be consistent, it indeed enforces that local symmetries, such as supersymmetry and gauge invariance, be realized both at the classical and at the quantum level  (via the cancellation of anomalies, see e.g.~\cite{IRU}).

It has recently been shown~\cite{CFI,VZD} that the conditions coming from the {\it localized BI} of eq.~(\ref{eq:FW}) are equally important. When both fluxes and D-branes are present, the bulk BI alone are not enough to ensure the consistency of the effective theory. The way localized BI enforce gauge invariance in the effective theory is multiple. For example, in $N=1$ type-IIA compactifications to four dimensions with non-trivial 3-form flux, the superpotential $W$ receives a contribution of the type (see e.g.~\cite{WHIIA,VZ2A}) 
\beq
W \quad \supset \quad - i \,  \int_{X_6} H \wedge \Omega^{c}
\quad \supset \quad  \int_{X_6} H \wedge C^{(3)}\,,\nn
\eeq
where $\Omega^c$ is the ``complexified" complex structure, whose  imaginary part is the RR 3-form potential $C^{(3)}$, and $X_6$ is the compact six-dimensional internal space. This superpotential term lifts some of the shift symmetries associated to the axions $C^{(3)}$. On the other hand, a D6-brane wrapping a cycle $\pi_6$ induces a St\"uckelberg gauging of the form
\beq
D_\mu C^{(3)} = \de_\mu C^{(3)}-[\pi_6]\, A_\mu  \,, \nn
\eeq
where $A$ is the U(1) gauge field living on the D6-brane. This means that under the U(1)
of the D6-brane the axions transform as
\beq
\delta_\lambda C^{(3)} = [\pi_6] \lambda\, . \nn
\eeq
For the effective action to be consistent, we must require the gauge invariance of the superpotential\footnote{Strictly speaking, the superpotential and the K\"ahler potential could be invariant up to a K\"ahler transformation, but this subtlety does not play a r\^ole in the present considerations.}, 
\beq
\delta_\lambda W \propto \int_{X_6} H \wedge \delta_\lambda C^{(3)}=
\lambda\int_{X_6} H \wedge [\pi_6]=\lambda\, \int_{\gamma_3} H =0 \,,
\qquad \gamma_3= X_6 \cap \pi_6 \,,\nn
\eeq
which is realized by the FW condition (\ref{eq:FW}). Analogously, in the presence of fluxes the cancellation of gauge anomalies in the effective theory  is realized only when {\it both} bulk and localized BI are satisfied \cite{VZD}. Finally, it was shown in \cite{KT}  that the condition (\ref{eq:FW}) is also crucial for the compatibility of non-perturbative effects and flux-induced gaugings. Indeed, in $N=2$ compactifications there would be a clash between Euclidean D2-brane instantons ($\pi_2$), which break the shift symmetry of $C^{(3)}$ via terms~\cite{BBS} such as
\beq
\exp\l( -\int_{\pi_2} \l[\Re (i \, e^{-\Phi} \Omega)+i C^{(3)}\r]\r) \,, \nn
\eeq
and $N=2$ gaugings induced by the $\ov{H}$ flux:
\beq
(G^{(4)})^2=(d C^{(3)}+ \ov{H} \wedge C^{(1)})^2\,. \nn
\eeq
Again, the condition (\ref{eq:FW}) prevents this clash from happening. Analogous mechanisms can be shown to hold also in type-IIB compactifications and for different D-brane setups.

We can now ask whether condition (\ref{eq:FW}) is actually the only one or there are more, in view of the fact that string dualities mix different kind of fluxes with themselves and with geometry itself, as well as
different kinds of branes. Unfortunately, from the mathematical point of view K-theory does not seem to fit 
with S-duality \cite{DMW} (see also \cite{Evslinrev} and references therein). This could mean either that K-theory is not enough to classify brane charges with general setups of fluxes, or that S-dualities do not hold exactly. We will study the interplay between string dualities and localized BI from the effective field theory point of view, showing the existence of more constraints relating localized sources and fluxes, in agreement with S-duality. We start by discussing the T-duality case, where results in this direction already exist, then move to S-duality. 

\subsection{T~duality}

Under T-duality, a D$p$-brane is mapped into a D$(p+1)$-brane or D$(p-1)$-brane,  depending on whether the dualized direction is orthogonal or parallel to the brane, respectively (when it is oblique the brane is mapped into a magnetized D$(p+1)$-brane). Analogously, the RR $p$-form potentials are mapped into $(p\pm1)$-form potentials. Finally, the NSNS sector is mapped into itself, with the metric mixing with the 2-form potential $B$, as a result of the interchange between KK and winding modes (see \cite{B}). As shown in \cite{KSTT}, if the T-dualized direction $k$ is parallel to the flux $H_{ijk}$, this is mapped into a twist of the geometry $\omega_{ij}^{\ \ k}$, which is equivalent to a vacuum expectation value for  the spin connection---a Scherk--Schwarz or geometric flux~\cite{SS}. On toroidal compactifications, the geometric flux $\omega$ can be easily taken into account by modifying the external derivative with a torsion term, \ie\ by making the replacement
\beq
d\to d+\omega \, . \nn
\eeq
In the absence of KK-monopoles, whose effects have been recently discussed in \cite{VZKK}, $\omega$ satisfies the consistency constraint \cite{SS}:
\beq 
\label{eq:omom}
\ov{\omega} \, \ov{\omega} \equiv \ov{\omega}_{[ij}^{\ \ k} \ov{\omega}_{l]k}^{\ \ m} =0\,. 
\eeq
When both $\omega$ and $H$ fluxes are considered, the effective external derivative becomes
\beq 
\label{eq:defD}
d_H\to \D =d+\omega+H\wedge\,,
\eeq
which is closed under the considered T-duality. However, if we keep using T-dualities, we eventually end up dealing with some `non-geometric' compactifications (for a recent review and references, see e.g. \cite{wecht}). We could generalize the external derivative also to these cases, to have a completely T-duality-invariant external derivative, but we will restrict ourselves here to the geometric case, where the effective field theory is better under control. Because of the condition (\ref{eq:omom}), the closure of the modified external derivative
\beq \label{eq:DD}
\D\D
=0\,, 
\eeq 
follows by imposing the modified NSNS BI
\beq
(d+\omega) H=0\,. \nn
\eeq
Analogously, the RR BI now read:
\beq
\label{eq:tBIs}
\D G= Q_R \, ,
\eeq
i.e.:
\beq 
\label{eq:tBI}
(d+\omega) G^{(8-p)}+ H\wedge G^{(6-p)}=
\sum_q Q_{p}(\pi_q) \, .
\eeq
By applying again $\D$ on eq.~(\ref{eq:tBI}), and using eq.~(\ref{eq:DD}), we then get the modified 
localized BI
\beq 
\label{eq:tFW}
\D ([\pi] e^\cF)=0 \,,
\eeq
which after expanding gives
\beq
(d+\omega)[\pi_p] = 0  \, ,
\qquad
\label{eq:SSpi}
(d+\omega )\cF+H = 0 \,,
\eeq
or, equivalently,
$$
\ov{\omega} \, [\pi_p] = 0 \, , 
\qquad
\ov{H} + \ov{\omega} \, \ov{\cF} = 0 \, ,
$$
where we momentarily included also the contributions from magnetic fluxes ($\ov{\cF}$).
As shown in \cite{VZD} (see also \cite{M}), the geometric flux modification in the first equation takes into account the fact that twisting the boundary conditions in the compactification changes the topology of the manifold,  in particular it removes some of the allowed cycles that cannot be used anymore to wrap branes on them. The second modification is instead the usual modification of the BI due to Scherk--Schwarz fluxes.
 
We can repeat the whole analysis of the compatibility between $\omega$ fluxes and D-branes, 
to find the same qualitative results as for the $H$ fluxes---in all cases the conditions (\ref{eq:tFW}) are crucial for the effective theory to be consistent, see e.g.~\cite{VZD}. T-duality thus mixes $H$ fluxes with torsion in the geometry, providing a geometric interpretation of the FW condition in the dual description. Roughly speaking, stable D$p$-branes must wrap manifolds that are non-trivial with respect to the `twisted' cohomology defined by the modified external derivative $\D$ of eq.~(\ref{eq:defD}).

\subsection{S duality}

We now turn to S-duality. This duality acts more dramatically on fields and sources, mixing the RR and NSNS sectors. The type-IIB theory is self-dual under S-duality: RR fields either are self-dual or are mapped into NSNS fields, and vice-versa. The same happens for branes: D-branes either are self-dual or are mapped into NSNS sources (fundamental strings and NS5-branes). In the type-IIA theory, S-duality controls the uplift to M-theory, where again the RR and NSNS sectors of type-IIA are mixed/unified: the 10D metric, the RR 1-form and the dilaton come from the 11D metric, while the RR 3-form and the NSNS 2-form come from the 11D 3-form. The condition (\ref{eq:tFW}), on the other hand, is not invariant under S-duality. This suggests that there should be also another condition involving RR-fluxes and NSNS-sources. We give here a short argument supporting the existence of this new condition and postpone the full derivation to section~\ref{sec:BI}, after discussing its M-theory origin in the next section. 

To get a non-trivial condition on NSNS sources, we need to include them in the BI,
\bea
(d+\omega)H&=&Q_{H}=\sum [\nu_5]\,, \nn \\
(d+\omega)\omega&=& Q_{KK}=\sum [\kappa_5] \, , 
\nn
\eea
where $\nu_5$ is the world-volume of NS5 sources and $\kappa_5$ of KK5-monopole sources~\cite{VZKK}. Because of this, our external derivative operator is no longer closed,
\beq
\label{eq:qNS}
\D \D = Q_{NS}\, ,
\eeq
where $Q_{NS}=Q_{KK}+Q_H$ is the generic NSNS source term. From here we can understand the problems in defining a cohomology with such external derivative.  If we repeat now the derivation of the condition (\ref{eq:tFW}) in the presence of NSNS sources, by acting on eq.~(\ref{eq:tBIs}) with the operator $\D$ we get, using also eq.~(\ref{eq:qNS}):
\beq
\D\D G = Q_{NS}\wedge G = \D Q_R\,. \nn
\eeq
In the absence of RR sources (and more generally for free NSNS sources) we have
\beq \label{eq:gBI}
[\nu_5]\wedge \ov G=0\qquad \Leftrightarrow \qquad 
\int_{\gamma_p \subset \nu_5} G^{(p)} = 0 \,,
\eeq
and 
\beq \label{eq:gBIK}
[\kappa_5]^\xi \ \ov G_\xi=0\qquad \Leftrightarrow \qquad 
\int_{\gamma_p \subset \kappa_5} G_\xi^{(p+1)} = 0 \,,
\eeq
where $\xi$ is the fibered circle of the KK5-monopole (see the next section) and the contraction of indices in eq.~(\ref{eq:gBIK}) reads
\beq
[\kappa_5]^\xi \ \ov G_\xi \equiv \frac{1}{3!(p-1)!}[\kappa_5]^\xi_{i_1 i_2 i_3} \ \ov G^{(p)}_{\xi \, j_1\dots j_{p-1}} 
dx^{i_1}\wedge dx^{i_2}\wedge dx^{i_3}\wedge dx^{j_1}\wedge \cdots \wedge dx^{j_{p-1}}\,. 
\eeq  
In the type-IIB theory, and for the RR 3-form field strength, the condition of eq.~(\ref{eq:gBI}) looks indeed like the S-dual of eq.~(\ref{eq:FW}), with NSNS and RR, fluxes and sources interchanged.

In the next section we will show how eqs.~(\ref{eq:FW}), (\ref{eq:gBI}) and (\ref{eq:gBIK}) have a common geometrical origin in M-theory, and how indeed they are connected via string dualities. In section~\ref{sec:MMS} we will extend the analysis to non-free branes, discussing flux-induced instabilities. Finally, in section~\ref{sec:BI} we will give a different derivation of the various conditions along the lines of the argument sketched above.

\section{Localized BI from M-theory}

Besides describing the strongly coupled limit of the type-IIA theory, M-theory gives a more geometric interpretation and an explicit `unification' of the two distinct sectors of the theory (RR and NSNS). As we will show below, M-theory also helps explaining the origin of the localized BI discussed before, providing the proof that the conjectured S-dual condition in eq.~(\ref{eq:gBI}) must also hold.

D-branes in type-IIA arise from brane-like solitonic objects in M-theory. In particular, D2-branes come from the dimensional reduction of M2-branes (the sources electrically coupled to the M-theory 3-form $A^{(3)}$). D4-branes arise from M5-branes (the magnetic duals of the M2-branes) wrapping the 11th dimension, which is taken to shrink to zero size in the weak string coupling limit. D0 and D6-branes arise instead as electric and magnetic sources of the eleven-dimensional (11D) metric.

Before discussing how conditions (\ref{eq:FW}), (\ref{eq:gBI}) and (\ref{eq:gBIK}) are connected via S- and T-dualities, and how M-theory unifies them, it is useful to review the M-theory origin of the by now well-known localized BI of eq.~(\ref{eq:FW}).

\subsection{D6-branes}
\label{sec:LBID6M}

In this section we consider the case of D6-branes in the type-IIA theory, which have a geometrical origin in M-theory, and show explicitly how the corresponding localized BI arise directly from the 11D M-theory effective action. We start from the following solitonic solution for the 11D metric of M-theory~\cite{T}:
\beq
ds^2=-dt^2 + \sum_{m=1}^6(dx^{m})^2 +ds^2_{\rm TNUT}\,, \label{eq:MD6metric}
\eeq
where
\beq
ds^2_{\rm TNUT}=f^{-1}(r)\l[dr^2+r^2d\theta^2+r^2\sin^2\theta d\phi^2\r]+f(r)\l[dx^{11}+V \r]^2\,, \nn
\eeq
and
\beq
f(r)=\l(1+\frac{m}{r}\r)^{-1}\,, \qquad V=m(1-\cos \theta)d\phi\,.  \nn
\eeq
The metric (\ref{eq:MD6metric}) describes a Taub--NUT geometry for a (6+1)-dimensional object where 
one of the four transverse dimensions is a circle ($x^{11}\sim x^{11}+4\pi m$). In the limit in which the size of this circle shrinks to zero, this geometry describes a (6+1)-dimensional object in 10D, sourcing a magnetic flux for the 11D graviphoton (from which the name KK6-monopole):
\beq 
\label{eq:defdV}
dV = m \, \sin \theta \,  d\theta \wedge d\phi\,.
\eeq
Since $V$ maps into the RR 1-form potential in the type-IIA limit,  the (6+1)-dimensional object is indeed a D6-brane. How the DBI action for the D6-brane can be recovered from the compactification of M-theory on the Taub--NUT geometry is discussed in \cite{BJO,I,S}, where it is also shown that the vector fields living on D6-branes arise from the M-theory 3-form potential. 

A fast way to show how eq.~(\ref{eq:FW}) arises from this compactification is the following. The $H$ flux of the type-IIA theory is generated in M-theory by the flux of the 4-form field strength $F^{(4)}$, with one leg along the 11th direction. We want to show that the flux component $F^{(4)}_{11mnr}$, where $(m,n,r)$ are along the D6-brane world-volume, is not compatible with the metric~(\ref{eq:MD6metric}). In the case under consideration, the background value for $F^{(4)}$ on (\ref{eq:MD6metric}) reads
\beq
F^{(4)}=H_{mnr} \, (dx^{11}+V)\wedge dx^{m}\wedge dx^{n}\wedge dx^{r}\,, \nn
\eeq
where the last factor arises because the metric is not diagonal.  In the absence of M5-branes, the 11D 4-form field strength $F^{(4)}$ satisfies the BI
\beq \label{eq:F4BI}
dF^{(4)}=0\, .
\eeq
We now want to see what this condition is mapped into, when calculated over the background of eq.~(\ref{eq:MD6metric}). Since $dV\neq 0$,
\beq
dF^{(4)}= H_{mnr} \, dV\wedge dx^{m}\wedge dx^{n}\wedge dx^{r}=0\,, \nn
\eeq
corresponds to requiring that $H$ vanishes on the D6-brane world-volume, i.e. to eq.~(\ref{eq:FW}).

To make contact with the discussion of the localized BI given in the previous section, we give now another way of obtaining this condition, which allows us to derive the full set of BI for the localized vector fields, and to highlight the connection with gauge invariance. To do this, we follow the method of \cite{I} for deriving the effective DBI action. It is useful to rewrite the 11D metric in a diagonal form,
\beq
ds^2=
-dt^2 + \sum_{m=1}^6(dx^{m})^2 
+f^{-1}(r) \l [ (\xi^{(1)}-C^{(1)})^2 +\tilde g_{ab} dx^a dx^b \r]\, , \nn
\eeq
where we kept the dependence on the fields associated to the fluctuations of the metric in the internal space, in particular the $3 \times 3$ block $\tilde{g}_{ab}$, whose background value is the flat metric, and the graviphoton $C^{(1)}$, and we introduced the modified 11th direction  
\beq
\xi^{(1)} = c \, \frac{f(r)}{m}\l[dx^{11}+m(1-\cos\theta)d\phi\r]\,, \nn
\eeq
which is actually an harmonic 1-form, with $c$ a normalization constant. As was the case for $V$, such a form is not closed, and its external derivative gives a self-dual 2-form
\beq
\xi^{(2)}=d\xi^{(1)}=\star_4\, \xi^{(2)}=\frac{f'(r)}{f(r)} dr \wedge \xi^{(1)}+c \, f(r) \, \sin\theta \, d\theta \wedge d\phi\,, \nn
\eeq
which is also harmonic and extends both in the $(r,\xi^{(1)})$ and in the $(\theta,\phi)$ directions. The second term is actually proportional to the magnetic flux ($dV$)  coming out of the KK6-monopole. In the type-IIA limit, this flux actually corresponds to the RR 2-form flux $G^{(2)}$ coming out of the D6-brane and satisfying the BI
\beq \label{eq:defpi6}
dG^{(2)}=[\pi_6]\,,
\eeq
which identifies the localized 3-form $[\pi_6]$ Poincar\'e dual to the D6-brane world-volume. The condition $d[\pi_6]=0$ directly follows from eq.~(\ref{eq:defpi6}).

It is convenient to expand $F^{(4)}$ over harmonic forms, in particular we can write:
\beq
F^{(4)} = G^{(4)}+\xi^{(1)} \wedge H +\xi^{(2)}\wedge\cF^{(2)}+\dots
\eeq
where we kept just the components that we will need in the following. As in \cite{I}, in the background (\ref{eq:MD6metric}), $\xi^{(1)}$ plays the role of the `11th' dimension, so that the different polarizations of $F^{(4)}$ can be identified with the  type-IIA RR 4-form $G^{(4)}$, the NSNS 3-form $H$ and the `localized' 2-form gauge fields $\cF$ on the world-volume of the D6-brane. Eq.~(\ref{eq:F4BI}) then becomes:
\beq 
\label{eq:expdF42}
dF^{(4)} = dG^{(4)}-\xi^{(1)}\wedge dH +\xi^{(2)}\wedge (d\cF^{(2)}+H)+\dots =0\,,
\eeq
i.e.
\beq
dG^{(4)}=0\,, \qquad 
dH^{(3)}=0\,, \qquad
d\cF^{(2)}+H=0\,   . 
\label{eq:BIfromF4}
\eeq

The first condition in eq.~(\ref{eq:BIfromF4}) is the RR BI for $G^{(4)}$ in the absence of D4-branes and with trivial $G^{(2)}$. The non-triviality of $G^{(2)}$ arising from eq.~(\ref{eq:defpi6}) has been reabsorbed into $\xi^{(2)}$  in this frame, \ie\ the $H$ term in the third equation. A non-trivial $G^{(2)}=dC^{(1)}$ can be implemented by considering also the 11D metric fluctuations around the background, since $C^{(1)}_M \propto g_{11 M}$. A topologically non-trivial $\ov G^{(2)}$, on the other hand, may arise by considering twisted tori where the Scherk--Schwarz parameter $\ov \omega_{ab}^{\ \ 11}$ is different from zero \cite{DP}: it would be interesting to check the BI conditions also in the presence of a non-vanishing 11D geometrical flux.

The second condition in eq.~(\ref{eq:BIfromF4}) is the NSNS BI in the absence of NS5-branes, which admits  the general solution
\beq
H=d B+\ov H\, . \nn
\eeq

Finally, the last condition in eq.~(\ref{eq:BIfromF4}) is the desired localized BI.  The solution of this BI reads
\beq \label{eq:sollBI}
\cF=dA+\ov F - B=F-B\,,
\eeq
provided that $\ov H=0$ (on the D6-brane world-volume), \ie\ the FW anomaly constraint be satisfied:
\beq
\ov H\wedge [\pi_6]=0\qquad \Leftrightarrow \qquad
\int_{\gamma_3 \subset \pi_6} H=0  \,. \nn
\eeq
This derivation makes it  clear why, in the DBI+WZ D6-brane action, the fields $B$ and $F$  always appear in the combination (\ref{eq:sollBI}), which is gauge invariant under the following transformations:
\beq
B\to B+d\lambda^{(1)} \,, \qquad A\to A+\lambda^{(1)}\,, \nn
\eeq
where $\lambda^{(1)}$ is a generic 1-form gauge transformation. Indeed, it is easy to show that the way the fields $B$ and $F$ transform under gauge transformations follows directly from the gauge transformations of the M-theory 3-form potentials. In the absence of fluxes we have
\bea
\delta A^{(3)}&=&d\Lambda^{(2)}\,,\nn \\
A^{(3)}&=&C^{(3)}-\xi^{(1)}\wedge B+\xi^{(2)} \wedge A +\dots\,,\nn \\
\Lambda^{(2)}&=&\lambda^{(2)}+ \xi^{(1)}\wedge\lambda^{(1)}+ \xi^{(2)}\,\lambda^{(0)} \,, \nn \\
\Rightarrow
\delta A^{(3)}&=&d\lambda^{(2)}-\xi^{(1)} \wedge d\lambda^{(1)}
	+ \xi^{(2)}\wedge (d\lambda^{(0)}+\lambda^{(1)}) \,,  \nn
\eea
hence
\beq \delta C^{(3)}=d\lambda^{(2)}\,, \qquad \delta B=d \lambda^{(1)}\,,\nn \eeq
\beq \delta A=d\lambda^{(0)}+\lambda^{(1)} \quad \Rightarrow \quad \delta F=d\lambda^{(1)}\,, \nn \eeq
so that the combinations $\cF=F-B$  is invariant.

We have thus shown that the gauge invariance of the D6-brane action follows from the gauge invariance associated with the M-theory 3-form potential. Moreover, the localized BI for D6-branes comes from the bulk M-theory BI for $F^{(4)}$. Imposing these BI on the Taub--NUT solution, the FW anomaly-cancellation constraint follows.

\subsection{KK5-monopoles}

Now that we have assessed the M-theory origin of the localized BI of eq.~(\ref{eq:FW}), we can try to see whether also eqs.~(\ref{eq:gBI}) and (\ref{eq:gBIK}) can be obtained in the same way. As already observed at the end of section~2 for the type-IIB case,  eq.~(\ref{eq:gBI}) is simply the S-dual of the better known FW anomaly-cancellation condition. Since in the type-IIA limit S-duality can be achieved by exchanging the role of the 11th dimension with one of the others, we can check what happens if we perform a different embedding of the Taub--NUT metric (\ref{eq:MD6metric}) in the 11D space-time. In particular, we can consider the KK6-monopole solution in 11D and identify the 11th dimension with one of its world-volume coordinates instead of the fibered circle of the Taub--NUT soliton. In this way, in the 10D  type-IIA limit, we get again a KK-monopole solution, which this time will have only five world-volume spatial dimensions, becoming a KK5-monopole. We can repeat the analysis of the previous section with the new assumption. The decomposition of the M-theory 4-form then reads:
\beq
F^{(4)}=G^{(4)}+\xi^{(1)}\wedge G^{(4)}_\xi +
 \eta \wedge H +  \xi^{(2)}\wedge\cK^{(2)}+\dots  \, , \nn
\eeq
where $\eta$ is the 1-form inside the KK6-monopole ($d\eta=0$) that plays the role of the 11th dimension, $G^{(4)}_\xi$ is the type-IIA RR 4-form with one leg on the $\xi^{(1)}$-monopole direction, and $\cK^{(2)}$ is the 2-form field strength localized on the IIA KK5-monopole~\cite{EL}.
Since $d\xi^{(1)}=\xi^{(2)}\neq0$, the BI for $F^{(4)}$,
\beq
d F^{(4)}=dG^{(4)}- \eta \wedge dH- \xi^{(1)} \wedge d G^{(4)}_\xi+ \xi^{(2)}\wedge (d\cK^{(3)}_\xi+G^{(4)}_\xi) +\dots=0\,,\nn
\eeq
gives a non-trivial condition on the KK5-monopole world-volume ($\kappa_5$)
\beq \label{eq:kk5BI}
[\kappa_5]^\xi \, \ov G^{(4)}_\xi=0\qquad \Leftrightarrow \qquad
\int_{\gamma_3  \subset \kappa_5} G^{(4)}_\xi = 0 \,. 
\eeq
This condition corresponds to eq.~(\ref{eq:gBIK}) and is the type-IIA `S-dual' of the FW anomaly-cancellation condition (\ref{eq:FW}). We can convince ourselves of this by T-dualizing the two conditions (\ref{eq:FW}) and (\ref{eq:kk5BI}) into the type-IIB theory, where the S-duality dictionary is more familiar. To do this, notice that in the type-IIA case S-duality has been achieved by exchanging $\eta$ and $\xi^{(1)}$ as 11th dimension: we thus want to use these directions for the T-dualization. We start with eq.~(\ref{eq:FW}), here $\xi^{(1)}$ was the 11th dimension that we had already shrunk, so we T-dualize along $\eta$, which in this case is a world-volume coordinate. If we choose $H$ orthogonal to $\eta$ it will stay invariant under T-duality, while the D6-brane is mapped into a D5-brane ($\pi_5$). Eq.~(\ref{eq:FW}) goes then into the condition:
\beq \label{eq:pi5FW}
\int_{\gamma_3\,\subset\, \pi_5} H=0\,. 
\eeq
In the case of eq.~(\ref{eq:kk5BI}), on the other hand, $\eta$ was the 11th direction. Then we T-dualize along $\xi^{(1)}$. $G^{(4)}_\xi$ maps into $G^{(3)}$, while, as discussed in \cite{BFRM, OV, BJO, Tong},  the KK5-monopole is mapped into a NS5-brane $\nu_5$. Therefore, under T-duality eq.~(\ref{eq:kk5BI}) goes into
\beq \label{eq:nu5FW}
\int_{\gamma_3\,\subset\, \nu_5} G^{(3)}=0\,,
\eeq
which is indeed the S-dual of eq.~(\ref{eq:pi5FW}) and agrees with eq.~(\ref{eq:gBI}). 
We may schematically summarize the result of the web of dualities as follows:
\begin{center}
\begin{tabular}{c c c c c c}
$\displaystyle \int_{\gamma_3\,\subset\, \kappa_6} F^{(4)}_{\xi}=0$ & $\displaystyle \xlongrightarrow{S^1_{(\xi)}}$ & 
$\displaystyle \int_{\gamma_3\,\subset\, \pi_6} H=0$ & $\displaystyle \xlongleftrightarrow{{\rm T_{\eta}}}$ 
& $\displaystyle \int_{\gamma_3\,\subset\, \pi_5} H=0$ &\\
M & &  IIA&  &  IIB & ${\displaystyle \Bigl )} \ {\rm S}$ \\
$\displaystyle \int_{\gamma_3\,\subset\, \kappa_6} F^{(4)}_{\xi}=0$ & $\displaystyle \xlongrightarrow{S^1_{(\eta)}}$ & 
$\displaystyle \int_{\gamma_3\,\subset\, \kappa_5} G^{(4)}_{\xi}=0$ & $\displaystyle \xlongleftrightarrow{{\rm T_{\xi}}}$ 
& $\displaystyle \int_{\gamma_3\,\subset\, \nu_5} G^{(3)}=0$ &
\end{tabular}
\end{center}
On the type-IIB side, also eq.~(\ref{eq:nu5FW}) can be understood as a localized BI. Indeed, by S-duality, as  fundamental strings ending on D5-branes produce a localized 2-form gauge field $\cF$ satisfying $d\cF+H=0$, D1-branes (-strings) ending on NS5-branes produce a localized 2-form gauge field $\cK^{(2)}$ satisfying $d\cK^{(2)}+G^{(3)}=0$.

\subsection{M5-branes}

A further example is represented by M5-branes, which may generate both D4-branes and NS5-branes in the type-IIA limit. Consider a M5-brane wrapping the manifold $\mu_5$ in the 11D space  ${\mathbb R}\times X_{10}$, where we take $X_{10}=X_{9}\times S^1$.  On the world-volume of the brane lives a self-dual 3-form field-strength $K^{(3)}$~\cite{BLNPST,APPS}, which always enters the M5-brane action in the gauge-invariant combination
\beq
{\cal K}^{(3)}=K^{(3)}-A^{(3)}\,. \nn
\eeq
This suggests that ${\cK}^{(3)}$ actually obeys the modified BI
\beq
d{\cal K}^{(3)}+F^{(4)}=0\,, \nn
\eeq
so that the following condition follows
\beq \label{eq:F4FW}
{\ov F}^{(4)}\wedge [\mu_5]=0\qquad \Leftrightarrow \qquad 
\int_{\gamma_4\,\subset\,\mu_5} F^{(4)}=0\,.
\eeq
After compactification, a non-trivial ${\ov F}^{(4)}$ may induce the gauging  of the shift symmetries of the axions coming from $A^{(6)}$, the 6-form dual to $A^{(3)}$, via the term
\beq 
\label{eq:F4gaug}
F^{(7)}=\de_\mu A^{(6)}+{\textstyle \frac12}{\ov F}^{(4)}\wedge A^{(3)}_\mu \,.
\eeq
The shift symmetries of the $A^{(6)}$ axions can however be broken by M5-instanton contributions
\beq
\sim \exp\l(i\int_{\mu_5} A^{(6)}\r)\,, \nn
\eeq
associated with Euclidean M5-branes. The condition (\ref{eq:F4FW}) then prevents the simultaneous presence  of the flux ${\ov F}^{(4)}$ and the M5-brane, when this would produce a clash with gauge invariance. The importance of eq.~(\ref{eq:F4FW}) to avoid inconsistencies between the gauging (\ref{eq:F4gaug}) and M5-instanton contributions has also been discussed nicely in \cite{AZ} for heterotic M-theory compactifications.

We can now perform the type-IIA limit in different ways. We can identify the 11th dimension with one of the directions parallel to $F^{(4)}$ and the M5-brane, in which case $F^{(4)}\to H$, $\mu_5\to \pi_4$ and $\cK^{(3)}\to\cF$, or with one of the orthogonal ones, in which case $F^{(4)}\to G^{(4)}$ and $\mu_5\to \nu_5$. So we get
\begin{center}
\begin{tabular}{c c c c c c}
$\displaystyle \int_{\gamma_3\,\subset\, \mu_5} F^{(4)}_{11}=0$ & $\displaystyle \xlongrightarrow{S^1}$ & 
$\displaystyle \int_{\gamma_3\,\subset\, \pi_4} H=0$ & $\displaystyle \xlongleftrightarrow{{\rm T_{x}}}$ & 
$\displaystyle \int_{\gamma_3\,\subset\, \pi_5} H=0$ &\\
M & &  IIA&  &  IIB & ${\displaystyle \Bigl )} \ {\rm S}$ \\
$\displaystyle \int_{\gamma_3\,\subset\, \mu_5} F^{(4)}_{\rm x}=0$ & $\displaystyle \xlongrightarrow{S^1}$ & 
$\displaystyle \int_{\gamma_3\,\subset\, \nu_5} G^{(4)}_{\rm x}=0$ & $\displaystyle \xlongleftrightarrow{{\rm T_{x}}}$ 
& $\displaystyle \int_{\gamma_3\,\subset\, \nu_5} G^{(3)}=0$ &
\end{tabular}
\end{center}
which shows once again that the conditions (\ref{eq:pi5FW}) and (\ref{eq:nu5FW}) have the same M-theory origin, which is indeed mapped into type-IIB S-duality.

From M-theory, we could also choose as 11th dimension a direction that is parallel to the M5-brane but
orthogonal to $F^{(4)}$. In this case the condition in the type-IIA limit would read
\beq \label{eq:pi4G4}
\int_{\gamma_4\,\subset\,\pi_4} G^{(4)}=0\,.
\eeq
Surprisingly, we have obtained a new constraint involving D-branes and RR-fluxes: we will discuss  this type of constraints further in section~\ref{sec:BI}.

A different duality web of constraints between fluxes and branes can be obtained starting from the condition that M5-branes wrap true cycles in a `twisted' M-theory compactification. Namely, we can require the condition
\beq \label{eq:SSM5}
\ov{\omega} [\mu_5]=0\,,
\eeq
where $\ov\omega$ is a geometrical flux in 11D. If $\ov\omega$ does not involve the 11th dimension, the condition (\ref{eq:SSM5}) maps into the analogous condition for D4-branes [eq.~(\ref{eq:SSpi})]. When instead the twist involves the 11th dimension, as argued before, $\ov \omega=\ov\omega_{ab}^{\ \ 11}$ maps into the RR 2-form flux $\ov G^{(2)}_{ab}$ of type-IIA. So we get the non-trivial set of conditions:
\begin{center}
\begin{tabular}{c c c c c c}
$\displaystyle \omega^{\rm x}[\mu_5]_{\rm x}=0$ & $\displaystyle \xlongrightarrow{S^1}$ & 
$\displaystyle \omega^{\rm x}[\pi_4]_{{\rm x}}=0$ & $\displaystyle \xlongleftrightarrow{{\rm T}_{{\rm x}}}$ & 
$\displaystyle \int_{\gamma_2\,\subset\, \pi_5} H_{\rm x}=0$ &\\
M & &  IIA&  &  IIB & ${\displaystyle \Bigl )} \ {\rm S}$ \\
$\displaystyle \omega^{11}[\mu_5]_{11}=0$ & $\displaystyle \xlongrightarrow{S^1}$ & 
$\displaystyle \int_{\gamma_2\,\subset\,\nu_5} G^{(2)}=0$ & $\displaystyle \xlongleftrightarrow{{\rm T}_{{\rm x}}}$ & 
$\displaystyle \int_{\gamma_2\,\subset\,\nu_5} G^{(3)}_{\rm x}=0$ &
\end{tabular}
\end{center}
Other analogous conditions can be obtained by exploiting different M-theory embeddings, M2-branes, S and T dualities. The interested reader should by now be able to work out the details.

\subsection{Consequences for the effective action}

As in the case of the FW constraint, each of the new conditions derived above translates into an important condition for the consistency of the effective action. This is somehow obvious from the fact that they can be related via T and S dualities to the usual FW constraint, or to the equivalent one [eq.~(\ref{eq:F4FW})] in M-theory, which ensure the consistency of the effective action.
We will give here a couple of examples. 

In the type-IIA theory, the condition
\beq \label{eq:G4FW}
\int_{\gamma_4\,\subset\,\nu_5}G^{(4)}=0\,, \nn
\eeq
forbids the existence of NS5-branes that wrap cycles with a non-trivial RR 4-form flux. Indeed the BI for the NSNS 7-form (Poincar\'e dual to the NSNS 3-form $H$)
\beq
dH^{(7)} + {\textstyle\frac12} G^{(4)} \wedge G^{(4)}=0\,, \nn
\eeq
tells us that in the presence of a non-trivial $\ov G^{(4)}$ flux the shift symmetry associated to the axion $B^{(6)}$ gets gauged in the effective 4D theory, since
\beq
H^{(7)}_{\mu abcdef}=\de_\mu B^{(6)}_{abcdef}+{\textstyle \frac12} \ov G^{(4)}_{[abcd} C^{(3)}_{ef]\mu}\,, \nn
\eeq
and this requires the effective potential to have an exact shift symmetry. However, Euclidean NS5-brane instantons can break in general  such a symmetry via terms like
\beq
\sim \exp\l(i\int_{\nu_5}B^{(6)}\r)\,.
\eeq
As already seen in the case of the usual FW condition, eq.~(\ref{eq:G4FW}) just prevents this inconsistency from happening. 

The common origin of the FW constraint and  its dual is probably more evident in the type-IIB theory, where S-duality is explicit. Consider an Euclidean D3-brane wrapping some 4-cycle: in general this will determine a breaking of the shift symmetries associated to the axion $C^{(4)}$, from the point of view of the lower-dimensional effective theory, via terms \cite{BBS}
\beq 
\label{eq:D3inst}
\sim\exp\l[\int_{\pi_3} \l( J\wedge J+i C^{(4)}\r)\r] \, .
\eeq 
On the other hand, the RR 5-form field strength reads
\beq
G^{(5)}_{\mu abcd}=\de_\mu C^{(4)}_{abcd}+{\textstyle \frac12} G^{(3)}_{[abc} B_{d]\mu}-{\textstyle \frac12} H_{[abc} C^{(2)}_{d]\mu}\,,
\eeq
therefore if either $H$ or $G^{(3)}$ (or both) are non-trivial on a 3-cycle, the axionic shift symmetry of $C^{(4)}$ is gauged. For this not to clash with eq.~(\ref{eq:D3inst}), we need a constraint between D3-branes and fluxes. One comes from the FW condition,
\beq
\int_{\gamma_3  \subset  \pi_3} H=0\,,
\eeq
while the other is its S-dual version:
\beq
\int_{\gamma_3 \subset  \pi_3} G^{(3)}=0 \, .
\eeq
We clearly need both of them not to run into inconsistencies. The latter constraints can easily be obtained via T-duality from the type-IIA constraint in eq.~(\ref{eq:pi4G4}). The general derivation will be given in section~\ref{sec:BI} [see eq.~(\ref{eq:nuBIonS})].

\section{Brane instabilities and non-free branes}
\label{sec:MMS}

In the previous sections we focused our attention on {\it free} brane configurations and on the consistency constraints arising from the BI for such configurations. As noticed in \cite{MMS}, however, the FW anomaly-cancellation condition can be relaxed in the presence of branes ending on the ``anomalous" brane. In particular, a D$p$-brane may support a non-trivial $H$-flux on its world-volume if a number of D($p-2$)-branes proportional to $H$ end on the D$p$-brane. This is because each D($p-2$)-brane behaves like a monopole source for the localized BI \cite{Str,Tbis}. When the ``anomalous" brane is localized in time, the process corresponds to the decay of N D$(p-2)$-branes into N units of $H$-flux via an instantonic D$p$-brane supporting the NS flux. This process \cite{MMS}  is also called MMS-instanton decay. In general, if $\pi_{(p-2)}$ is the space wrapped by a D($p-2$)-brane,  and $\ov H$ a NS flux orthogonal to the brane (namely $[\pi_{(p-2)}]\wedge \ov H=0$), then if there exists a Euclidean D$p$-brane supporting $\ov H$ and containing the D$(p-2)$-brane itself, the latter may be unstable against MMS-instanton decay. Notice that such condition corresponds to requiring that
\beq \label{eq:MMS1}
[\pi_{(p-2)}]=\ov H \wedge [\pi_{p}]=d_H [\pi_{p}]\,,
\eeq
\ie\ that the D$(p-2)$-brane does  not wrap a trivial cycle [see eq.~(\ref{eq:stab1})] in the twisted cohomology constructed with the external derivative operator $d_H$. It is easy to guess the answer when also geometrical fluxes are added (see also \cite{LB}), namely
\beq
[\pi_{(p-2)}]= \ov{\omega} [\pi'_{(p-2)}]+ \ov{H} \wedge [\pi_{p}]=\D \sum_q Q_{{}_{(p-2)}} (\pi_q)\,.
\eeq

Analogously, for the FW condition we can derive the constraint (\ref{eq:MMS1}) from M-theory. Consider again the M-theory embedding of a D6-brane, namely a KK6-monopole, and perform the reduction of the BI for $F^{(4)}$ on a Taub--NUT background, this time in the presence of a M5-brane. The 11D BI reads
\beq \label{eq:BIF4M5}
dF^{(4)}=[\mu_5]\, .
\eeq
The decomposition of the 4-form in terms of harmonic forms of the Taub--NUT background was given in section~\ref{sec:LBID6M}, we can do the same for the 11D 5-form $[\mu_5]$ dual to the M5-brane,
\beq \label{eq:decmu5}
[\mu_5]=[\pi_4]-\xi^{(1)} \wedge [\nu_5]+\xi^{(2)} \wedge [\pi_4^\perp]_{\theta \phi}+\dots \, ,
\eeq
where as before we keep only the relevant components. The first term in the decomposition arises when all five indices of $[\mu_5]$ are in the KK6 world-volume: this means that the M5-brane wraps the 11th dimension, so that from the 10D point of view it is a D4-brane $\pi_4$. The second term, on the other hand, does not wrap the 11th dimension and corresponds to a NS5-brane in type-IIA. For the last term, finally, notice that $\xi^{(2)}$ contains the factor $dV$ of eq.~(\ref{eq:defdV}), so that the M5-brane extends in the radial direction $r$ and in the fibered direction $\xi^{(1)}$ of the monopole: this corresponds to a D4-brane ending into the D6-brane.
If we now look at the reduction of the BI (\ref{eq:BIF4M5}) and project over the basis of forms, we get
\beq
d G^{(4)}=[\pi_4]\,, \qquad
dH=[\nu_5]\,,\qquad
d\cF+H=[\pi_4^\perp]_{\theta \phi}\, .
\eeq
As expected, the RR and NSNS BI get localized contributions from the corresponding sources. As anticipated, also the localized BI on the D6-brane gets a localized contribution, from D4-branes ending on the D6-brane. The integrability condition of this BI is thus modified: a D6-brane can support N units of $H$-flux, as long as there are N D4-branes ending on it\footnote{Actually, the $\xi^{(2)}$ form in eq.~(\ref{eq:decmu5}) has also a $(r,\xi)$ component, which corresponds to a M5-brane extending along the $(\theta,\phi)$ directions, \ie\ a NS5-brane wrapping the 2-sphere surrounding the monopole. From the 10D point of view its contribution to the localized BI corresponds to the fact that the NS5-brane screens the D6-brane from the $H$ flux. Since there is nothing new in this configuration we will neglect this component of $\xi^{(2)}$ here and in the following.}.

Analogously to the derivation of the generalized FW conditions from M-theory, we can now perform the type-IIA limit on eq.~(\ref{eq:BIF4M5}), by identifying as 11th dimension a different direction in the world-volume of the KK6-monopole. If we do this, the l.h.s. of eq.~(\ref{eq:BIF4M5}) is expanded as in eq.~(\ref{eq:expdF42}), while $[\mu_5]$ reads:
\beq
[\mu_5]=[\pi_4]-\eta \wedge [\nu_5]-\xi^{(1)} \wedge [\pi_4]_\xi+ \xi_{(2)}\wedge [\pi_4^\perp]_{\theta \phi}+\dots,
\eeq
where $[\pi_4]$ and $[\nu_5]$ correspond to a D4-brane and a NS5-brane wrapping  the fibered circle of the KK5-monopole, $[\pi_4]_\xi$ is a D4-brane localized on the $\xi^{(1)}$ direction and, as in the previous example, $[\pi_4^\perp]_{\theta \phi}$ is a D4-brane that ends on the tip of the KK5-monopole.
So, finally, the 11D BI gives
\beq
dG^{(4)}=[\pi_4]\,,\ dG^{(4)}_\xi=[\pi_4]_\xi\,, \qquad
dH=[\nu_5]\,, \qquad d\cK^{(2)}+G^{(4)}_\xi=[\pi_4^\perp]_{\theta \phi}\,.
\eeq
Besides the usual BI for RR and NSNS fields, we get also a modification for the BI of the 3-form localized on the KK5-monopole, coming from D4-branes ending on the KK5-monopole. Therefore the constraint of eq.~(\ref{eq:kk5BI}) can be relaxed, i.e. a KK5-monopole can support a $G^{(4)}_{\xi}$-flux with one leg on the fibered circle, if D4-branes that end on the monopole are added. This also means that D4-branes may be unstable in the presence of a RR-flux, via a MMS-like instanton process involving a KK5-monopole. This process has a nice geometrical interpretation: the decaying D4-brane wraps the 1-cycle $\xi^{(1)}$, which is the fibered circle of the KK5-monopole; because of this the cycle is trivial, the D4-brane may unwind passing through the tip of the monopole \cite{GHM} and shrink to zero size. What is left is just flux by conservation of RR charge.

We can now T-dualize our brane and flux setup along either of the two directions $\eta$ and $\xi^{(1)}$. The result for the integrability conditions is, schematically:
\begin{center}
\begin{tabular}{c c c c c c}
$\displaystyle dF^{(4)}=[\mu_5]$ & $\displaystyle \xlongrightarrow{S^1_\xi}$ & 
$\displaystyle \int_{\gamma_3\subset \pi_6} H=N^\perp_{D4}$ & $\displaystyle \xlongleftrightarrow{{\rm T_{\eta}}}$ 
& $\displaystyle \int_{\gamma_3\subset \pi_5}  H=N^\perp_{D3}$ &\\
M & &  IIA&  &  IIB & ${\displaystyle \Bigl )} \ {\rm S}$ \\
$\displaystyle dF^{(4)}=[\mu_5]$ & $\displaystyle \xlongrightarrow{S^1_\eta}$ & 
$\displaystyle \int_{\gamma_3\subset\kappa_5} G^{(4)}_\xi=N^\perp_{D4}$ & 
$\displaystyle \xlongleftrightarrow{{\rm T_{\xi}}}$ & $\displaystyle \int_{\gamma_3\subset\nu_5} G^{(3)}=N^\perp_{D3}$ &
\end{tabular}
\end{center}
where $N^\perp_{Dp}$ stands for the number of D$p$-branes ending on the `anomalous' brane. 
Therefore also for the MMS-instanton process we get its S-dual version. Since together with the FW condition the MMS decay can be seen as defining some cohomology (actually a K-theory), the fact that we get for both conditions an S-dual version seems to suggest that also for the RR sector there must be
some sort of cohomological/geometrical interpretation, although at the moment we ignore in which form.

We can extend the discussion to other configurations of fluxes and branes and find analogous constraints and decay processes. In section~\ref{sec:BI} we will present the general method to derive the constraints directly from the bulk BI in 10D theories, which may be used to get the relevant conditions in a generic setup. 

\subsection{MMS instantons and interpolating domain walls}

MMS-like configurations play an important r\^ole when dealing with domain walls interpolating between two string vacua (see e.g.~\cite{EM}). Consider the 10D space to be the product of the four-dimensional space-time and a compact six-manifold, i.e. ${\mathbb R}^{(1,3)} \times X_6$. A D$p$-brane that extends along time and two out of the three non-compact directions is a domain wall from the 4D point of view. Such brane will then wrap a $(p-2)$-cycle of $X_6$ and will source a flux for the RR $G^{(8-p)}$ field strength. At distances larger than the typical size of $X_6$, the flux will be directed towards the non-compact direction orthogonal to the domain wall, so that the latter will produce a jump of the $G^{(8-p)}$ flux, polarized along the $(8-p)$-cycle dual to the $(p-2)$-cycle of the brane,  between the two regions divided by the wall. The domain wall can thus interpolate between two string vacua with different $G^{(8-p)}$ flux, the difference being determined by the D$p$-brane tension.

If the $G^{(8-p)}$ flux is not constrained by BI this is the end of story. In this way, for instance, we can interpolate between the different type-IIA $AdS_4$ vacua of \cite{VZ2A} with different  values of the moduli by considering D2 or D4 (wrapped on a 2-cycle of $X_6$) domain walls, which make the unconstrained $G^{(6)}$ and $G^{(4)}$ fluxes jump.
 
A more interesting situation arises when the domain wall sources a flux that is constrained by tadpole cancellation conditions. In this case the domain wall cannot just source a flux, it must also source branes! Indeed, since the tadpole cancellation condition relates the number of branes to the units of flux, if the latter jump across the wall, the number of branes must also jump in order to preserve the BI on both sides of the wall. This phenomenon is enforced by imposing also localized BI. We illustrate this case with an example. 

Consider the type-IIA vacua introduced before. The configuration of branes and fluxes will in general satisfy the RR BI:
\beq \label{eq:BIG2}
d G^{(2)}+\omega G^{(2)}+H G^{(0)}=\sum_{D6/O6} [\pi_6]\,.
\eeq
%
\EPSFIGURE[t]{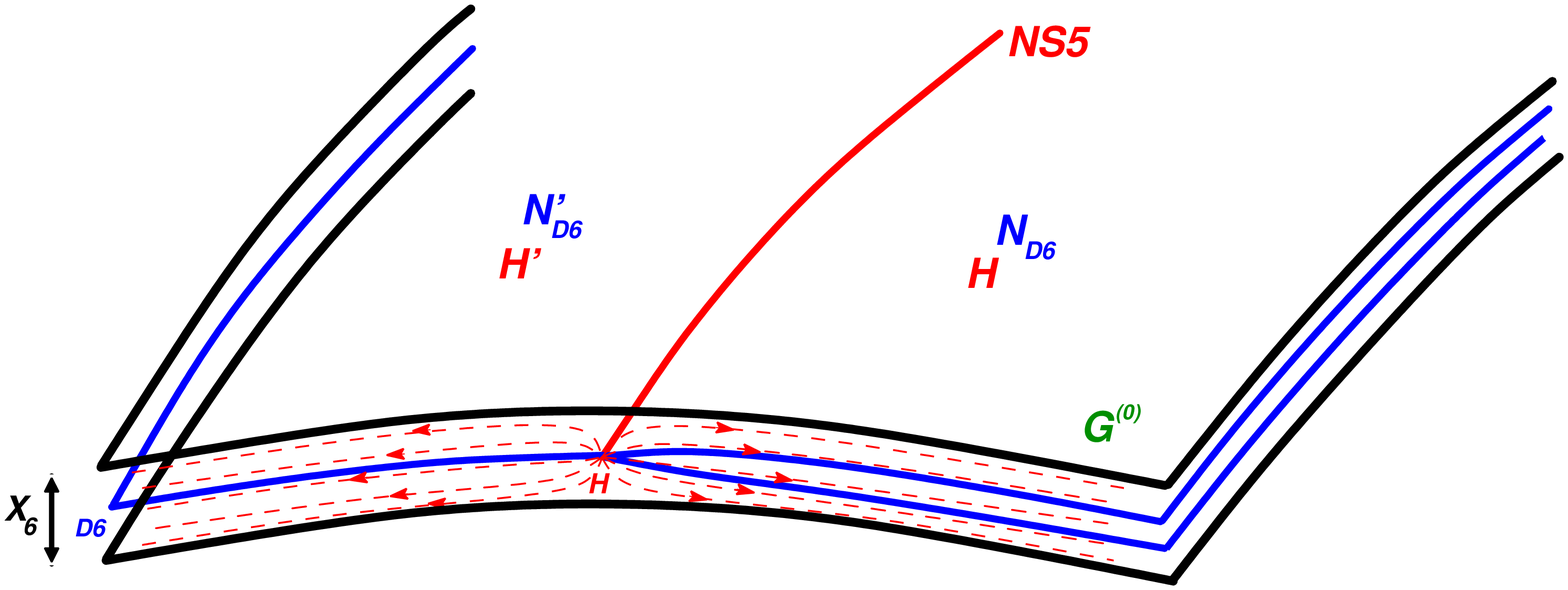,width=1\textwidth}{\label{fig:interpol} NS5 domain wall interpolating between two vacua of 
massive type-IIA. The RR BI in eq.~(\ref{eq:BIG2}) requires that
the discontinuity of the NSNS flux produced by the NS5 ($\Delta H=H-H'=Q_{NS5}$) be compensated by a discontinuity in
the number of D6-branes $\Delta N_{D6}=N_{D6}-N'_{D6}=G^{(0)}\Delta H$. Therefore $\Delta N_{D6}=Q_{NS5} G^{(0)}$ must
end on the NS5-branes as required by the cancellation of the `anomaly' (\ref{eq:gBI}), see also eq.~(\ref{eq:N5GNp}).}
A D8-brane that wraps the whole $X_6$ behaves in 4D as a domain wall that makes the flux $\ov G^{(0)}$ 
jump by one unit. This means that the number of D6-branes must jump by $H$-units in order to preserve eq.~(\ref{eq:BIG2}). Notice however that a D8-brane wrapping the whole $X_6$ will suffer a FW anomaly because of the $H$-flux. For the anomaly to be cancelled, $H$ D6-branes must end on the D8-brane, making eq.~(\ref{eq:BIG2}) consistent on both sides of the wall. Analogously, we can use a D6-brane wall, sourcing a jump of $G^{(2)}$ flux, but failing to satisfy the closure condition (\ref{eq:SSpi}), and require $\omega$ D6-branes to end on it \cite{LB}.

The same arguments can now be applied for NS5 and KK5 domain walls, which originate jumps of the  $H$ and $\omega$ fluxes, respectively: these are inconsistent in the presence of RR-fluxes, unless D-branes are allowed to end on the domain walls (see e.g. fig.~\ref{fig:interpol}).

Analogously in type-IIB, D5 and NS5 domain walls, which source a jump in the $G^{(3)}$ and $H$ fluxes respectively, may be used to interpolate between different IIB vacua. An explicit example already exists \cite{KPV}.
Since these branes are anomalous in the presence of non trivial $H$ and $G^{(3)}$ fluxes on their worldvolume, D3-branes must end on them. The actual number of D3-branes needed to restore consistency of the BI is just the right one to satisfy the RR tadpole cancellation condition on the vacua on both side of the wall, namely
\beq
\ov H\wedge \ov G^{(3)}=\sum_{D3/O3} [\pi_3]\,.
\eeq

The connection between flux-constraints, bulk BI and branes ending on branes can be made clearer by re-deriving the various conditions with another method that we will now discuss in the next section.
 
\section{General constraints from bulk BI}
\label{sec:BI}

In this section we present a full derivation of the various constraints, both for free branes (FW-like) and for non-free branes (MMS-like), which uses the bulk BI of the 10D effective actions, implementing the method sketched in section~\ref{sec:rev}.

A D$p$-brane in 10D is a codimension-$(9-p)$ object that sources a flux for the RR $G^{(8-p)}$ 
form on the $(8-p)$-sphere surrounding it\footnote{For simplicity, we will restrict our discussion
to the case in which branes are away from singularities, such as orbifold fixed points and orientifold
planes.}. We consider now the BI for $G^{(10-p)}$,
\beq \label{eq:10mpBI}
dG^{(10-p)}+H\wedge G^{(8-p)}= \sum_q Q_{(p-2)} (\pi_q) \, , 
\eeq
calculated on a $(11-p)$-dimensional closed manifold, $S^{(8-p)}\times \gamma_3$, defined as the product of the  $(8-p)$-sphere supporting the flux and a generic 3-cycle $\gamma_3$ (if it exists) in the world-volume of the D$p$-brane. If the brane is isolated, we can always choose the $S^{(8-p)}$ so that it does not intersect any other object. In particular, in the absence of D$(p-2)$-brane charges, the BI (\ref{eq:10mpBI}) gives the condition
\beq \label{eq:BIonS}
0=\int_{S^{(8-p)}\times \gamma_3} \l(dG^{(10-p)}+H\wedge G^{(8-p)}\r) \propto \int_{\gamma_3} H \,,
\eeq
which corresponds to the well-known FW anomaly-cancellation condition of eq.~(\ref{eq:FW}). 

Exactly the same steps can be performed in the case of a NS5-brane, to derive the condition of eq.~(\ref{eq:gBI}). In this case we have an $H$ flux on the $S^3$ surrounding the NS5-brane. We can then calculate the BI (\ref{eq:10mpBI}) on $S^3 \times \gamma_{(8-p)}$, where $\gamma_{(8-p)}$ is a $(8-p)$-cycle in the NS5 world-volume, and get
\beq
0=\int_{S^{3}\times \gamma_{(8-p)}} \l(dG^{(10-p)}+H\wedge G^{(8-p)}\r) \propto \int_{\gamma_{(8-p)}} G^{(8-p)}\,, \nn
\eeq
\beq
\Rightarrow [\nu_5]\wedge \ov G=0\,, \nn
\eeq
which is exactly eq.~(\ref{eq:gBI}).

Finally, in the case of the KK5-monopole we can surround the monopole with an $S^2$ that supports the geometrical flux $\omega^{\xi}\propto dV^{\xi}$. Because of the geometrical flux, the BI (\ref{eq:10mpBI}) receives an extra torsion contribution. If we now integrate it over $S^2\times \gamma_{(7-p)}$, where $\gamma_{(7-p)}$ is a $(7-p)$-cycle in the KK5 world-volume, we get
\beq \label{eq:KBIonS}
0=\int_{S^{2}\times \gamma_{(7-p)}} \l(d G^{(8-p)}+\omega^\xi G^{(8-p)}_\xi+H\wedge G^{(6-p)}\r) 
\propto \int_{\gamma_{(7-p)}} G^{(8-p)}_\xi\,,
\eeq
\beq
\Rightarrow [\kappa_5]^\xi \, \ov G_\xi=0\,,
\eeq
which coincides with eq.~(\ref{eq:gBIK}).

Before turning to the case of branes ending on other branes, we discuss the last type of conditions involving RR fluxes and D$p$-branes, such as the one in eq.~(\ref{eq:pi4G4}). Indeed, there is another non-trivial BI we can use, the one associated to the NS 7-form $H^{(7)}$, magnetic dual of $H$:
\beq \label{eq:BIH7}
dH^{(7)}+{\textstyle \frac12}\sum_p G^{(p)}\wedge G^{(8-p)}=Q(\nu_1)\,,
\eeq
where $Q(\nu_1)$ is the charge associated to the fundamental string (NS1-brane). In the presence of a D$p$-brane we can integrate eq.~(\ref{eq:BIH7}) over $S^{(8-p)}\times \gamma_{p}$, where as before $S^{(8-p)}$ is a sphere surrounding the D-brane and $\gamma_p$ is a $p$-cycle in the brane world-volume. We thus get
\beq \label{eq:nuBIonS}
0=\int_{S^{(8-p)}\times \gamma_p} \l( dH^{(7)}+\frac12 \sum_q G^{(q)}\wedge G^{(8-q)}\r)=\int_{\gamma_p}G^{(p)}\,,
\eeq 
which in the special case $p=4$ gives eq.~(\ref{eq:pi4G4}). 

We can now discuss non-trivial configurations with branes ending on other branes. We discuss first the D$p$-brane case in detail. As will become clear in a moment, the other cases follow analogously. Basically, we want to repeat the step of the first example in this section in the presence of a D$(p-2)$-brane ending on the D$p$-brane. Eq.~(\ref{eq:BIonS}) will now get also a non-trivial contribution from D$(p-2)$-branes
\beq
N_{(p-2)}^{\perp}\equiv\int_{S^{(8-p)}\times \gamma_3}[\pi_{(p-2)}]=
\int_{S^{(8-p)}\times \gamma_3} \l(dG^{(10-p)}+H\wedge G^{(8-p)}\r) \propto \int_{\gamma_3} H \, .
\eeq
$N_{(p-2)}^{\perp}$ is indeed the intersection number of our D$(p-2)$-brane with $S^{(8-p)} \times\gamma_3$, which counts the number of D$(p-2)$-branes ending on the D$p$-brane minus the number of those leaving from the D$p$-brane that are orthogonal to the polarization of $H$, namely
\beq
N_p\,\int_{\gamma_3} H=N^{\perp}_{(p-2)}\,,
\eeq
where $N_p=\ov G^{(8-p)}\wedge [S^{(8-p)}]$ is the number of D$p$-branes sourcing the flux $\ov G^{(8-p)}$. 
Schematically the branes and the fluxes are embedded as follows
\begin{center}
\begin{tabular}{ccccccc}
$\gamma_3$ &$\times$& ${\mathbb R}^{p-2}$ &$\times$& ${\mathbb R}$ &$\times$& $S^{(8-p)}$ \\ \hline
D$p$ & & D$p$ & & & & $G^{(8-p)}$ \\
$H$ &&D$(p-2)$ && D$(p-2)$ &&
\end{tabular} \,.
\end{center}
We thus recovered eq.~(\ref{eq:MMS1}), i.e. the fact that a D$p$-brane may support N units of $H$-flux as long as N D$(p-2)$ branes end on it, or equivalently N D$(p-2)$-branes in the presence of N-units of (orthogonal) $H$ flux may decay into an instantonic D$p$-brane supporting the $H$ flux. The derivation could also be done by integrating the BI over $S'^{(8-p)}\times \gamma_3$, where $S'^{(8-p)}$ is the $(8-p)$-sphere surrounding the D$p$-brane without the point of intersection with the D$(p-2)$-brane. In this case there is no contribution from the $[\pi_{(p-2)}]$ term of the BI, but the manifold $S'^{(8-p)}\times \gamma_3$ has now a non-trivial boundary (see fig.~\ref{fig:MMS1}). It is easy to check that the term $dG^{(10-p)}$ gives the missing contribution
\beq
\int_{S'^{(8-p)}\times \gamma_3} dG^{(10-p)}=\int_{\de S'^{(8-p)}\times \gamma_3} G^{(10-p)}=N^\perp_{(p-2)}\,.
\eeq
%
\EPSFIGURE[t]{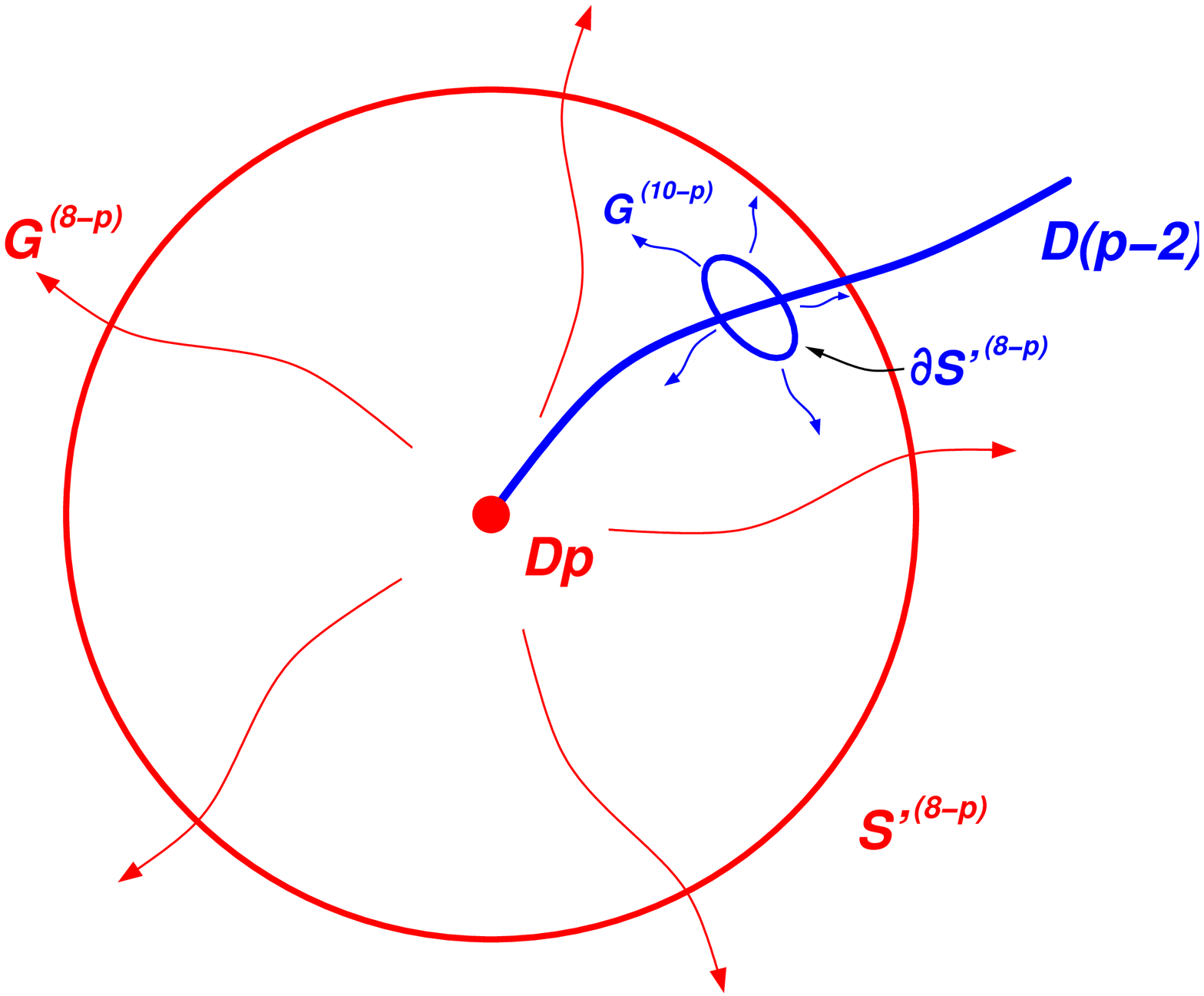,width=.6\textwidth}{\label{fig:MMS1} A D$(p-2)$-brane ends on a D$p$-brane supporting
a $H$ flux  on its worldvolume (not shown in the figure). The $(8-p)$-sphere surrounding the D$p$-brane 
and supporting its RR flux has a boundary with a non trivial RR flux ($G^{(10-p)}$) from the 
D$(p-2)$-brane that compensates for the `anomalous' contribution to the BI, as discussed in the text.}

The case with NS5-branes and RR-fluxes can be discussed in the same way. In this case we get the following consistency condition
\beq  \label{eq:N5GNp}
N_{5}\int_{\gamma_{(6-p)}} G^{(6-p)}=N_{p}^\perp \, ,
\eeq 
where now $N_5$ is the number of NS5-branes, $\gamma_{(6-p)}$ is a $(6-p)$-cycle contained in the NS5 and $N_p^\perp$ is the number of D$p$-branes ending on the NS5-branes and whose world-volumes have non-trivial intersection number with $\gamma_{(6-p)}$  (in other words,  the flux $G^{(6-p)}$ must be polarized orthogonally to the D$p$-brane). We thus get the striking result that a D$p$-brane can be unstable also in the presence of RR-fluxes. For example, in massive type-IIA compactifications ($\ov G^{(0)}\neq0$) D6-branes may decay into instantonic NS5-branes.
With a $G^{(2)}$ flux, D4-branes may decay via instantonic NS5-branes, and so on.

Analogously, for KK5-monopoles we must add the contribution from D$p$-branes ending on the monopole, and eq.~(\ref{eq:KBIonS}) becomes in this case
\beq
N_p^\perp\equiv\int_{S^{(2)}\times \gamma_{(7-p)}}[\pi_p]=\int_{S^{(2)}\times \gamma_{(7-p)}} 
\l(d G^{(8-p)}+\omega^\xi G^{(8-p)}_\xi+H\wedge G^{(6-p)}\r) 
\propto \int_{\gamma_{(7-p)}} G^{(8-p)}_\xi\,,
\eeq
where $N_p^\perp$ is the intersection number of D$p$-branes with $S^{(2)}\times \gamma_{(7-p)}$, i.e. the number of D$p$-branes ending on the monopole, wrapping the fibered circle $\xi$ and orthogonal to the flux $G^{(8-p)}_\xi$, which as before is taken to be polarized along the world-volume of the monopole (namely on $\gamma_{(7-p)}$). We thus have
\beq
N_K \int_{\gamma_{(7-p)}} G^{(8-p)}_\xi=N_p^\perp\,,
\eeq
where $N_K$ is the number of KK5. So we get that a KK5 may support a RR flux over its world-volume and the fibered circle as long as there are D-branes ending on it, and equivalently D$p$-branes may decay via a KK5 instanton if they wrap a circle that is shared by a RR flux which is otherwise polarized orthogonally to the brane. In this case it is possible to understand the process in geometrical terms: the circle wrapped by the D-brane gets trivialized by the KK5 instanton, the D-brane may unwrap the circle passing through the tip of the monopole/instanton as described in \cite{GHM}, the D-brane is thus allowed to shrink to zero size and annihilate, the RR flux guarantees the conservation of charge.

Finally, it is now easy to guess what happens with eq.~(\ref{eq:nuBIonS}): D$p$-branes may support N units of $G^{(p)}$ fluxes as long as N fundamental strings end on the D-brane, and analogously N fundamental strings may decay in the presence of N units of RR fluxes $G^{(p)}$ via Euclidean D$p$-brane instantons. This type of configurations have already been discussed in the context of the holographic correspondence in \cite{Wb}.

\section{Conclusions}

In string compactifications, the simultaneous presence of fluxes and localized sources is highly constrained by Bianchi identities. Besides the usual RR and NSNS tadpole cancellation conditions, corresponding to Gauss'  law for brane charges, additional localized constraints arise when fluxes and localized sources are simultaneously present. In this paper we derived some new localized constraints that can be interpreted as a generalization of the Freed--Witten anomaly-cancellation condition. We  showed that all these localized constraints are related by a web of string dualities and can be derived directly from M-theory compactifications, where their geometrical nature is more manifest. We discussed their importance for the consistency of the lower-dimensional effective theory, stressing that they enforce gauge invariance of the effective action when fluxes and branes are simultaneously present,  and that they may be relevant for protecting the flat directions, associated with the axionic symmetries gauged by fluxes, from instanton corrections arising from Euclidean branes. We also studied the possibility of relaxing these localized constraints in the presence of branes ending on other branes. This case can be relevant for the study of interpolating solutions in the landscape (see e.g.~\cite{EM,KLPT} for recent discussions and references to earlier work), since the latter allow to connect vacua with a different number of branes and a different content of fluxes. In analogy with the known case, where a FW-anomalous instantonic brane can trigger decays of D-branes in the presence of $H$ flux, we found that both D-branes and NS-branes may decay in the presence of NSNS as well as RR fluxes, which may be relevant for the study of the non-perturbative (in)stability of non-supersymmetric vacua.

There are a number of aspects that could deserve further study. For example,  we did not discuss the extension to non-trivial magnetic fluxes, which by affecting the RR charge of the brane may modify the localized BI, changing the structure of the local constraints and possibly allowing new decay processes. Another aspect that can be analyzed is the effect that these new consistency conditions have on actual string compactifications, in particular how strong is the constraint on the interplay of the various sources for moduli potentials, such as fluxes, branes and non-perturbative effects. We have shown indeed that the new localized BI are crucial for the consistency of the effective action when fluxes and branes are both present. We have also shown, however, that the constraints may be relaxed by just adding some other localized sources ending on the `anomalous' brane. It would be interesting to understand in detail how this translates into the lower-dimensional theory, whether for example it corresponds to adding some new degrees of freedom to the effective field theory that restore gauge invariance. Finally, it would be interesting to understand whether there is a general way of classifying all the possible constraints and allowed setups, since it would give an extra tool to characterize the underlying structure of the string landscape. 

\section*{Acknowledgments}
We thank Luca Martucci and Alessandro Tomasiello for helpful discussions.
GV thanks the Galileo Galilei Institute for Theoretical Physics for the hospitality and the INFN for partial support during the completion of this work.  FZ thanks the Laboratoire de Physique Th\'eorique de l'\'Ecole Normale Superieure, Paris, for its warm hospitality during the final phase of this work. This research was supported in part by the European Programme ``The Quest For Unification'', contract MRTN-CT-2004-503369.

\end{document}